\documentclass[%
superscriptaddress,
showpacs,preprintnumbers,
nofootinbib,
 amsmath,amssymb,
 aps,
prd,
notitlepage
]{revtex4-1}

\usepackage[english]{babel}
\usepackage[utf8x]{inputenc}
\usepackage[T1]{fontenc}

\usepackage{amsmath}
\usepackage{graphicx}
\usepackage[colorinlistoftodos]{todonotes}
\usepackage[colorlinks=true, allcolors=blue]{hyperref}

\begin{document}
\title{Relativistic stars in a cubic Galileon Universe}
\author{Hiromu~Ogawa}
\email[Email: ]{jh.ogawa"at"rikkyo.ac.jp}
\affiliation{Department of Physics, Rikkyo University, Toshima, Tokyo 171-8501, Japan
}
\affiliation{Institute of Cosmology and Gravitation, University of Portsmouth, Portsmouth, PO1 3FX, UK}

\author{Tsutomu~Kobayashi}
\email[Email: ]{tsutomu"at"rikkyo.ac.jp}
\affiliation{Department of Physics, Rikkyo University, Toshima, Tokyo 171-8501, Japan
}
\author{Kazuya~Koyama}
\email[Email: ]{kazuya.koyama"at"port.ac.uk}
\affiliation{Institute of Cosmology and Gravitation, University of Portsmouth, Portsmouth, PO1 3FX, UK}

\begin{abstract}
We study relativistic stars in Hordenski theories that evade the gravitational wave constraints and exhibit the Vainshtein mechanism, focusing on a model based on the cubic Galileon Lagrangian. We derive the scalar field profile for static spherically symmetric objects in asymptotically de Sitter space-time with a linear time dependence. The exterior solution matches to the black hole solution found in the literature. Due to the Vainshtein mechanism, the stellar structure is indistinguishable from that of General Relativity with the same central density as long as the radius of the star is shorter than the Vainshtein radius. On the other hand, the scalar field is not suppressed beyond the Vainshtein radius. These solutions have an additional integration constant in addition to the mass of the star.  
\end{abstract}
\pacs{04.50.Kd}
\preprint{RUP-19-29}

\maketitle

\section{Introduction}
The recent detection of the gravitational wave event, GW170817, from a neutron star binary merger by LIGO and VIRGO~\cite{TheLIGOScientific:2017qsa} and its electromagnetic counterpart, the gamma-ray burst GRB170817A~\cite{Monitor:2017mdv,Savchenko:2017ffs,Goldstein:2017mmi}, placed a stringent constraint on the propagation speed of gravitational waves at low-redshift, $z\sim 0.008$. The result shows that the propagation speed of gravitational waves is close to that of light, more precisely $(c_T^2-c^2)/c^2\leq 6\times 10^{-15}$, where $c_T$ is the speed of gravitational waves and $c$ is the speed of light. This result places tight constraints on a large class of modified gravity theories, in particular the Horndeski theory, which is the most general scalar-tensor theory with second-order equations: two of the four free functions present in the Horndeski theory is strongly constrained~\cite{Lombriser:2015sxa, Lombriser:2016yzn,  Ezquiaga:2017ekz,Creminelli:2017sry,Sakstein:2017xjx,Baker:2017hug} (see Ref.~\cite{deRham:2018red} for a discussion of the validity of these constraints) \footnote{ We should note that this constraint only applies to the models in which the scalar field has non-negligible contributions to the cosmological background evolution today.}. The decay of gravitational waves further constrains the extension of the Horndeski theory \cite{Creminelli:2018xsv, Creminelli:2019nok}.  As a result, only the simplest models known as Kinetic Gravity Braiding~\cite{Deffayet:2010qz,Pujolas:2011he, Kobayashi:2010cm} survive, which is a generalization of the cubic Galileon.  The cubic Galileon models, in particular those that try to explain the late time acceleration without the cosmological constant, are strongly constrained by the galaxy-Integrated Sachs Wolfe effect cross correlation \cite{Renk:2017rzu,Peirone:2017vcq}. These models are also constrained by the Solar System constraint as well as astrophysical tests \cite{Sakstein:2017bws} \footnote{Recently, it was shown that the cubic Galileon also suffers from instabilities in the presence of gravitational waves and this puts strong constraints on the model \cite{Creminelli:2019kjy}.}. However, these models are the simplest Horndneski theories beyond k-essence to understand the strong gravity behaviour of these theories.
	
The direct detection of gravitational waves has opened a new window to test the theory of gravity in a strong gravity regime. The study of compact objects, such as black holes and neutron stars in the surviving scalar-tensor theory, allows us to examine the no-hair theorem that holds in General Relativity (GR) with a minimally coupled scalar field, and to investigate the efficiency of the screening mechanisms inside compact stars. In some models, even if the deviations from GR are small in vacuum, the scalar field can acquire a large value inside neutron stars, and the structural properties of neutron stars can be different from those in GR. This phenomenon is first pointed out in the context of the Brans-Dicke theory.~\cite{Damour:1993hw,Damour:1996ke}. The equation of state independent relations~\cite{Yagi:2013bca,Hinderer:2009ca} break
the degeneracy between the uncertainties in the equation of state of nuclear matter and the effects associated with the scalar-tensor theory. Compact objects thus provide a natural laboratory to test gravity in a strong gravity regime.

Nevertheless, in shift-symmetric Horndeski theories, the no-hair theorem holds for black holes~\cite{Hui:2012qt} and neutron stars~\cite{Barausse:2015wia, Barausse:2017gip, Lehebel:2017fag}. 
The no-hair theorem was established under the assumption that the scalar field has a static configuration. Recently, it was pointed out that if the scalar field depends on time linearly,  $\phi(t,r)=qt+\Psi (r)$, a non-trivial scalar field configuration can be found for black holes (see \textit{e.g.}~ \cite{Babichev:2013cya,Kobayashi:2014eva,Babichev:2016fbg}, and \cite{Herdeiro:2015waa,Babichev:2016rlq} for a review of black hole solutions with a scalar hair) and relativistic stars (see for example~\cite{Maselli:2016gxk} for neutron star solutions in the fab-four theory) in shift-symmetric Horndeski theory. Very recently, relativistic stars have been studied in beyond Horndeski theory~\cite{Babichev:2016jom, Sakstein:2016oel} and in DHOST theory~\cite{Kobayashi:2018xvr}. These theories exhibit a partial breaking of the Vainshtein mechanism inside a star~\cite{Kobayashi:2014ida,Koyama:2015oma,Saito:2015fza,Crisostomi:2017lbg, Langlois:2017dyl,Dima:2017pwp}. Thus, the structure of a star was significantly modified in these theories while restoring GR in the solar system. 

The purpose of the present work is to study relativistic stars embedded in a de-Sitter Universe in the cubic Galileon theory. Unlike previous studies where analytic black hole solutions can be found with a linearly time dependent scalar field, black hole solutions can be found only numerically in this theory~\cite{Babichev:2016fbg}. These solutions are embedded in a de-Sitter spacetime and have three branches of solutions according to their effective cosmological constant. It is not clear whether these black hole solutions can be an exterior solution of a relativistic star. In addition, in this theory, the Vainshtein mechanism works inside a matter source at least in the weak gravity limit. It was shown that this is also the case for relativistic stars \cite{Chagoya:2014fza} but until now, no fully relativistic star solution was constructed in this theory. In this paper, we will construct a relativistic star solution numerically and study the structure of the star as well as the exterior solution. 

This paper is organised as follows. In section II, we present the cubic Galileon theory and introduce black hole solutions. In section III, we derive the equations to describe a relativistic star, which correspond to the Tolman-Oppenheimer-Volkoff equation in GR, and discuss the boundary conditions to solve these equations. In section IV, we present numerical solutions for a relativistic star with the polytropic equation of state and discuss the deviation of the stellar structure from GR and their asymptotic behaviours. Section V is devoted to conclusions. 

\section{Theory}
\subsection{Equation of motion}
Throughout this paper, we consider a cubic Galileon theory which is a
subclass of Kinetic Grabity Brading~\cite{Deffayet:2010qz,Pujolas:2011he}. 
The action of the cubic Galileon is given by
\begin{equation}
    S=\int d^4 x \sqrt{-g}[\zeta (R-2\Lambda)-\eta (\partial \phi)^2+\gamma \Box \phi (\partial \phi)^2+\mathcal{L}_{\textrm{m}}],\label{eq:action3}
\end{equation}
where $\zeta, \eta$ and $\gamma$ are constant parameters, $\Lambda$ is the bare cosmological constant and $\mathcal{L}_{\textrm{m}}$ is the matter Lagrangian. 
%The action is often considered in the literature (see, \textit{ e.g.,}\cite{Babichev:2016fbg,Dvali:2000hr, Deffayet:2010qz,Babichev:2012re}). 
The equations of motion deriving from (\ref{eq:action3}) are given by
\begin{align}
    &\zeta(G_{\mu\nu}+\Lambda g_{\mu\nu})+ H_{\mu\nu}=\frac{1}{2} T_{\mu\nu},\\
    &\nabla_{\mu}J^{\mu}=0,\label{eq:scalarfield}\\
    &\nabla_{\mu}T^{\mu\nu}=0,
\end{align}
where
\begin{align}
   & H_{\mu\nu}=\eta \left[\dfrac{1}{2} g_{\mu \nu} (\partial \phi)^2 - \partial_\mu \phi \: \partial_\nu \phi \right] 
 -\gamma \left[- \Box \phi \: \partial_\mu \phi \: \partial_\nu \phi + \partial_{(\mu} \phi \: \partial_{\nu)} (\partial \phi)^2 \vphantom{\left[(\partial \phi)^2\right]} 
-\dfrac{1}{2} g_{\mu \nu} \partial^\rho \phi \: \partial_\rho \left[(\partial \phi)^2\right] \right], \\
&T_{\mu\nu}=(\rho +P)u_{\mu}u_{\nu}+Pg_{\mu\nu}, \label{eq:conserveq}\\
&J^{\mu}=\partial_\nu \phi \left[g^{\mu\nu}(\gamma \Box \phi-\eta) -\gamma \nabla^\mu \nabla^\nu \phi \right],
\end{align}
$u$ is the unit 4-velocity of a perfect fluid with the energy
density $\rho$ and pressure $P$. The conserved current $J^{\mu}$ is defined as $\sqrt{-g}J^{\mu}=\delta S/\delta (\partial_{\mu}\phi)$.

In this paper, we consider a static and spherically symmetric spacetime
\begin{equation}
    ds^2=-h(r)dt^2+\frac{dr^2}{f(r)}+r^2d\Omega^2.\label{eq:lineelement}
\end{equation}
The action (\ref{eq:action3}) possesses a shift symmetry $\phi \rightarrow \phi +\textrm{const.}$, and $\phi$ does not appear without its derivatives in the equations of motion. Therefore, the scalar filed can be dependent on time linearly as shown in \cite{Babichev:2013cya,Babichev:2016fbg}
\begin{equation}
    \phi(r,t)=qt + \int dr \frac{\chi(r)}{h(r)},\label{eq:phian}
\end{equation}
where $q$ is a constant parameter. This ansatz for the scalar field was used to obtain black hole and neutron star solutions in scalar-tensor theory, since the ansatz breaks the assumptions for the no-hair theorems for black holes and neutron stars in shift-symmetric Horndeski theories~\cite{Babichev:2013cya,Kobayashi:2014eva,Babichev:2016rlq,Babichev:2016fbg,Maselli:2016gxk} (and see,\textit{e.g.,}~\cite{Herdeiro:2015waa,Babichev:2016rlq} for a review). 

\subsection{Black hole solutions}
Let us summarize the black hole solutions obtained in
\cite{Babichev:2016fbg}. The asymptotic behaviour of the black hole solutions is described by the homogeneous solutions. The homogeneous cosmological solutions in this theory are given by~\cite{Babichev:2016fbg}
\begin{align}
     &f(r)=h(r)=1-\frac{\Lambda_{\textrm{eff}}}{3}r^2,\label{eq:fh}\\
     &\chi(r)=\frac{\eta r}{3\gamma}, \label{eq:chi}
\end{align}
with an effective cosmological constant $\Lambda_{\textrm{eff}}$:
\begin{equation}
\Lambda_\mathrm{eff}=
\left\{
\begin{array}{r c l}
\Lambda_< &=& \dfrac{1}{2} \left(\Lambda+ \sqrt{\Lambda^2+4\: \Lambda_\mathrm{KGB}^2}\right) \ \text{if} \ \eta <0,\\
\\
\Lambda_>^+ &=& \dfrac{1}{2} \left(\Lambda + \sqrt{\Lambda^2-4\: \Lambda_\mathrm{KGB}^2}\right) \ \text{if} \ \eta >0 \ \text{and} \ |\Lambda | > 2\: \Lambda_\mathrm{KGB},\\
\\
\Lambda_>^- &=& \dfrac{1}{2} \left(\Lambda- \sqrt{\Lambda^2-4\: \Lambda_\mathrm{KGB}^2}\right) \ \text{if} \ \eta >0 \ \text{and} \ |\Lambda| > 2\: \Lambda_\mathrm{KGB},
\end{array}
\right.
\label{eq:Lambdaeff4D}
\end{equation}
where 
\begin{equation}
\Lambda_{\textrm{KGB}}=\left(\frac{|\eta|^3}{6 \zeta \gamma^2} \right)^{1/2}.
\end{equation}

In FLRW coordinates, the solutions are written as 
$$
ds^2 = - \mathrm{d}\tau^2 + e^{2 H \tau} (d \rho^2 + \rho^2 d\Omega^2),
$$
and the scalar field is homogeneous $\phi(\tau,\rho)=q_0 \tau$ where $q_0$ is given by 
\begin{equation}
q_0^\pm \equiv \left[\dfrac{\zeta \Lambda}{\eta} \pm \sqrt{\left(\dfrac{\zeta \Lambda}{\eta} \right)^2 - \dfrac{2 \eta \zeta}{3 \gamma^2}}\; \right]^{1/2}.
\label{eq:q}
\end{equation}

One can find black hole solutions (\ref{eq:lineelement}) and (\ref{eq:phian}) numerically that are asymptotically de Sitter spacetime 
~(\ref{eq:fh}-\ref{eq:q}) with a range of parameters and the scalar field velocity $q$. Numerical solutions with $\Lambda_{<}$ and $\Lambda_{>}^+$ were obtained in \cite{Babichev:2016fbg}. The solution is characterized by $q$, which determines the time dependence of the scalar field in (\ref{eq:phian}). Note that $q$ can be different from the cosmological value $q_0$. 
%Even if $q \neq q_0$, the scalar field is still asymptotically homogeneous. 
Since $q$ is not determined by the parameters of the model, it can be considered as a scalar hair if it is different from $q_0$. In \cite{Babichev:2016fbg}, solutions with $q \neq q_0$ were found.  

\section{Tolman-Oppenheimer-Volkoff equation}

\subsection{Equations of motion}
Using the ansatz (\ref{eq:lineelement}) and (\ref{eq:phian}), we obtain the equations of motion for the scalar field $\phi$ and the metric variables $f$ and $h$ as follows:
\begin{align}
    &\gamma q \left(r^{4}h\right)' \frac{f}{h} \chi^{2}
    -\gamma q^{3} r^{4} h' - 2 \eta q r^4 h \chi=0, \label{eq:tr} \\
    &\eta r^{2} \left(\frac{f}{h}\chi^{2} - q^{2} \right) + 2 \zeta r f h' + 2 \zeta h \left(-1+f + \Lambda r^{2}\right)-r^2h P = 0, 
    \label{eq:rr}
    \\
    &2\zeta r^2h^2[-1+\Lambda r^2+(rf)']+\eta q^2r^4 h\left[1+\frac{f}{h}\left(\frac{\chi}{q}\right) \right]\notag\\
    &+\gamma q^3\left[-2r^4hf\left(\frac{1}{h}\frac{\chi}{q}\right)'+2r^4f^2y^2\left(\frac{1}{h}\frac{\chi}{q} \right)'-y(r^4f)'+r^4y^3\frac{f'}{h} \right]+r^4h^2\rho=0,
     \label{eq:tt}
\end{align}
where a prime denotes a derivative with respect to $r$.
Eqs.~(\ref{eq:tr}), (\ref{eq:rr}) and (\ref{eq:tt}) are the ($tr$),($rr$) and ($tt$) components of the Einstein equations, respectively. 
Note that for the shift-symmetric action, the $(tr)$ component of the metric equations is equivalent to $J^r=0$. This implies that the equations of motion for the scalar field (\ref{eq:scalarfield}) is automatically satisfied. From the conservation equation (\ref{eq:conserveq}), we obtain
\begin{equation}
    P'=-\frac{\rho +P}{2}\frac{h'}{h}.
    \label{eq:pressure}
\end{equation}
Finally, for the numerical calculations, we consider the polytropic equation of state
\begin{equation}
    \rho=\left(\frac{P}{K}\right)^{1/2}+P,\label{eq:eoseq1}
\end{equation}
with $K$ is a constant. This equation of state have been frequently considered in the context of modified gravity \cite{Kobayashi:2018xvr,Silva:2016smx,Maselli:2016gxk,Cisterna:2015yla,Cisterna:2016vdx}. We finally obtained the set of the equations, (\ref{eq:tr}), (\ref{eq:rr}), (\ref{eq:tt}) and \eqref{eq:pressure}, which correspond to the  Tolman-Oppenheimer-Volkoff equation in GR. 

To solve the field equations numerically, following the approach in \cite{Babichev:2016fbg}, we introduce a mass scale $\bar{m}$ and define the dimensionless radius, pressure, density, and three dimensionless constants as combination of the parameters of theory as follows:
\begin{align}
    &\bar{r}=\bar{m}r,\;\; \bar{\rho}=\frac{\rho}{2 \bar{m}^2 \zeta},\;\; \bar{P}=\frac{P}{2 \bar{m}^2 \zeta},\notag\\
    &\alpha_1=-\frac{\gamma q \bar{m}}{\eta},\;\;
    \alpha_2=-\frac{\eta q^2}{\zeta \bar{m}^2},\;\;
    \alpha_3=\frac{\Lambda}{\bar{m}^2}.
\end{align}
Using these dimensionless quantities, one can rewrite the field equations as follows:
\begin{align}
    &\alpha_{ 1 }\left(\bar{r}^{4}h\right)'\frac{f}{h}y^{2}+2\bar{r}^{4}hy
    -\alpha_{ 1 }\bar{r}^{4}h'= 0,\\
    &\alpha_{2}\bar{r}^{2}\left[1-\frac{f}{h}y^{2}\right]+2h\left(-1+f+\alpha_{3}\bar{r}^{2}\right)+2\bar{r}fh'-2\bar{r}^2h\bar{P}=0,\\
    &2\bar{r}^2h^2(-1+\alpha_3\bar{r}^2+(\bar{r}f)')-\alpha_2\bar{r}^4h\left(1+\frac{f}{h}y^{2}\right)\notag\\
    &+\alpha_1\alpha_2\left[-2\bar{r}^4hf\left(\frac{y}{h}\right)'+2\bar{r}^4f^2y^2\left(\frac{y}{h} \right)'-y(\bar{r}^4f)'+\bar{r}^4y^3\frac{f'}{h} \right]+2r^4h^2\bar{\rho}=0.
\end{align}
where $y(r) = \chi(r)/q$. Hereafter, we will omit the bar from $r$ for simplicity. In our numerical simulations, $\bar{m}^{-1}$ is chosen to be roughly the radius of the star. 

\subsection{Boundary conditions}
In order to integrate these equations, we require boundary conditions at the centre of the star. To derive the boundary conditions at the centre of the
star, we expand $f$, $h$, $\chi$, $\rho$, and $P$ in the following forms:
\begin{align}
    &h(r)=h_c+\frac{h_2}{2}r^2+\cdots,\;\; f(r)=1+\frac{f_2}{2}r^2\notag\\
    &\chi(r)=\chi_cr+\cdots \\
    &\rho(r)=\rho_c+\frac{\rho_2}{2}r^2,\cdots\;\; P(r)=P_c+\frac{P_2}{2}r^2\cdots\notag
\end{align}
where $h_2$, $f_2$, $\rho_2$, and $P_2$ are constants, and $h_c$, $\rho_c$ and $P_c$ are central values. These constants can be found in terms of $h_c$ and $\rho_c$.

We numerically solve above equations in the following manner:
we impose the boundary conditions at the centre of the star for a set of parameters $\alpha_1, \alpha_2$ and $\alpha_3$. Given $h_c$ and $\rho_c$, we can integrate the equations from the centre until reaching the surface of
the star $r=R$, where the pressure vanishes $P(R)=0$.
Then, we match the interior solution to the vacuum solution, which asymptotes the de-Sitter solution
\begin{equation}
\begin{split}
     &h(r)\sim f(r)\sim -C_1 r^2 \;\; \mathrm{at} \; r\rightarrow \infty,\\
      &\chi(r)\sim -C_2 r \;\;\mathrm{at} \; r\rightarrow \infty,
\end{split}
\label{eq:outer}
\end{equation}
with positive constants $C_1$ and $C_2$. We tune the central value $h_c$ so that the exterior solution satisfies (\ref{eq:outer}). The central density $\rho_c$ determines the mass of the star. 

\section{Numerical Results}

\subsection{Stellar structure }
Fig.1 shows an example of the solutions for $h(r), f(r)$ and $y(r)$ in the interior of the star. We found that, in the interior, the solutions for $h(x)$, $f(x)$, $\rho(r)$ and $P(r)$ are indistinguishable from the GR solution with the same central density $\rho_c$ in this example. This confirms that the Vainshtein mechanism is working even in the strong gravity regime in this theory. 

\begin{figure}[h]
	\centering
	\includegraphics[width=15cm]{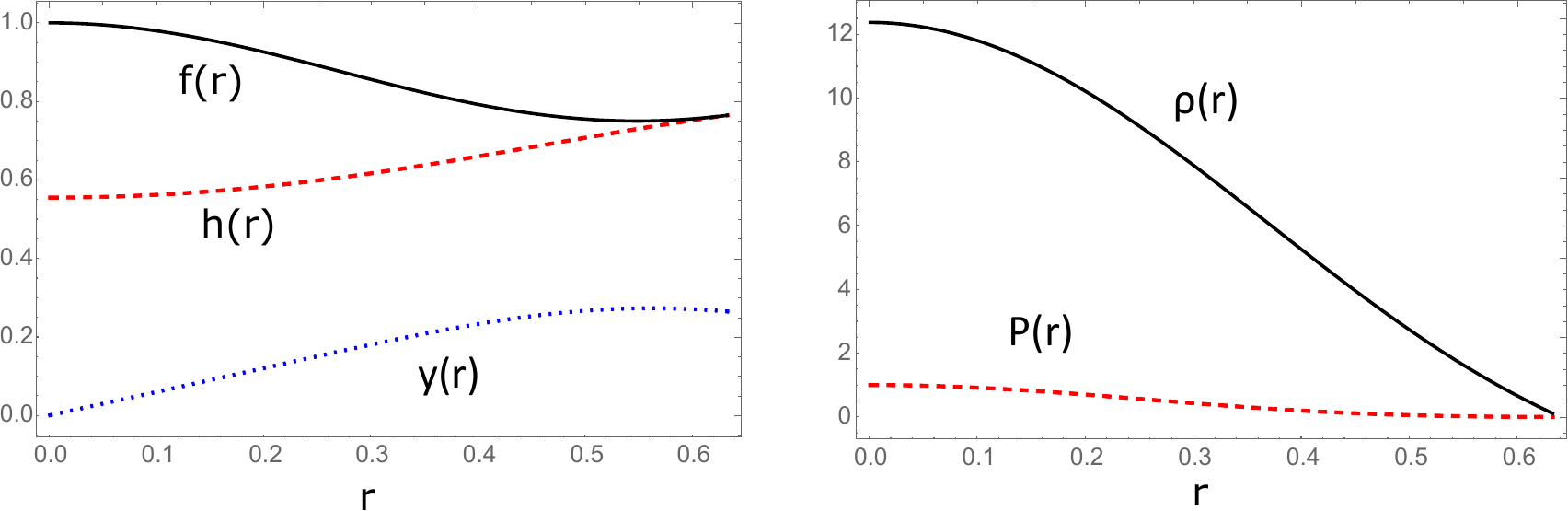}
	\caption{The behaviours of metric functions and the scalar field inside the star with $\alpha_2=-2.5 \times  10^{-7}$, $\alpha_3=10^{-4}$, $K=773 \times10^{-5}$ and $\rho_c=12.4$ in the $\Lambda_{\rm eff}=\Lambda^{+}_{>}$ branch. We choose $\alpha_1=\sqrt{2/3(\alpha_2+2 \alpha_3)}$, which gives $q=q_0$. These solutions are not distinguishable from the solutions in GR with the same central density of the star. The surface of the star is given by $R=0.63$.}
	\label{Fig1}
\end{figure}

The Vainshtein radius below which the Vainshteim mechanism operates, can be estimated as follows in the weak gravity limit. 
At large distance but smaller than the cosmological horizon, the metric components $h$ and $f$ are written as follows
\begin{equation}
h=f=1-\frac{\mu}{r},\label{eq:2}
\end{equation}
where $\mu$ is a constant. Using the expression~(\ref{eq:2}), we can rewrite the scalar field equation (\ref{eq:tr}) as 
\begin{equation}
\alpha_1 \left(\frac{y}{r} \right)^2+\frac{1}{2} \left(\frac{y}{r} \right) -\frac{\alpha_1 \mu}{4r^3} = 0.
\end{equation}
At large distance, the linear term dominates and we obtain the solution as
\begin{equation}
\frac{y}{r}\sim \frac{\alpha_1 \mu}{r^3}.
\end{equation}
The non-linear term dominates when $y/r \sim 1/\alpha_1$. Therefore, we can estimate the Vainshtein radius below which the non-linear term dominates
\begin{equation}
r_{\rm v} \sim  (\alpha_1^2 \mu)^{1/3}.    
\end{equation}
Below the Vainshtein radius $r<r_{\rm v}$, the non-linear term suppresses the scalar field and the effect of the scalar field is negligible. For the parameters used in Fig.1, the Vainshtein radius is O(10) larger than the radius of the star, confirming that the Vainshtein mechanism is operating to suppress the deviations of GR in terms of the stellar structure. 

\subsection{Exterior solution}
We now look at the exterior solution. The asymptotic solutions at large $r$ are given by ~\cite{Babichev:2016fbg}
\begin{align}
h(r) & = - \frac{\Lambda_{\rm eff}}{3} r^2 + 1 - \frac{\mu}{r},
\label{larger1}
 \\
f(r) & = - \frac{\Lambda_{\rm eff}}{3} r^2 + c_f^{(0)} - \frac{\mu}{r}, 
\label{larger2}
\\
y(r) &= -\frac{1}{3 \alpha_1} r + \frac{c_{\chi}^{(-1)}}{r} - 
\frac{\mu c_{\chi}^{(-2)}}{r^2},
\label{larger3}
\end{align}
where $c_f^{(0)}, c_{\chi}^{(-1)}, c_{\chi}^{(-2)}$ are constants that are determined by the model parameters. For $q=q_0$, $c_f^{(0)}=1$ and $c_{\chi}^{(-1)} =0$. We checked that our numerical solutions asymptotes these solutions at large $r$, confirming that the exterior solution is the black hole solution found in \cite{Babichev:2016fbg}. At large $r$, the leading order terms dominate and we have $h(r)=f(r)$ and $y(r)$ is linearly proportional to $r$ (Fig.2). The scalar field sourced by the star outside the Vainshtein radius is not suppressed and the scalar charge is given by $q \mu c_{\chi}^{(-2)}$ as $y(r)=\chi(r)/q$. Note that $q$ is an additional integration constant to $\mu$ and it is not determined by the model parameters if $q \neq q_0$. In addition, $\mu$ for a given central density of the star depends on $q$ as shown in Fig 3 even though the stellar structure has no dependence on $q$ as long as the Vainshtein radius is larger than the size of the star. This is because the mass $\mu$ contains not only the contribution from the star but also the energy density of the scalar field. If $q\neq q_0$, the scalar field acquires an additional profile because of $c_{\chi}^{(-1)}  \neq 0$. 

\begin{figure}[h]
	\centering
	\includegraphics[width=15cm]{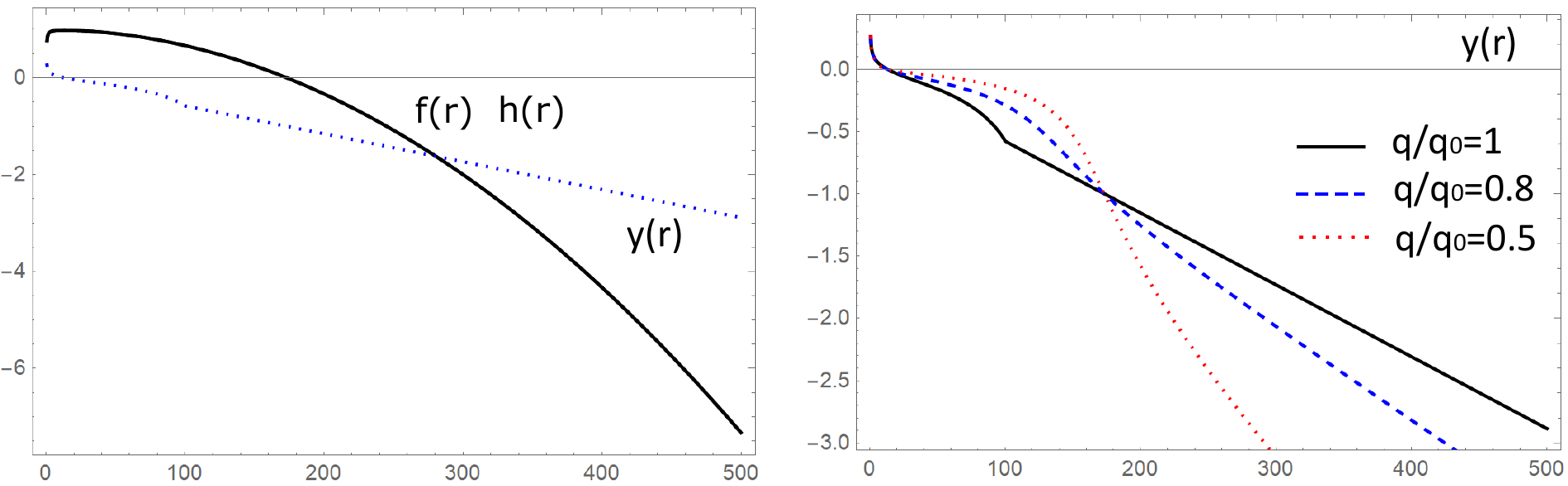}
		\label{Fig2}
	\caption{Left: The behaviours of metric functions and the scalar field outside the star with the same parameters as Fig.1. The cosmological horizon is at $r=173$. 
	Right: The exterior solution with $q/q_0=1, 0.8$ and $0.5$ where $\alpha_2=-2.5 \times  10^{-7} (q/q_0)^2$ and $\alpha_1=\sqrt{2/3(\alpha_2+2 \alpha_3)} (q/q_0)$. Note that the gradient of $y(r)$ appears to be discontinuous for $q/q_0=1$ at around $r=100$, but the derivatives of $y(r)$ are finite. This is due to the transition to the cosmological solution proportional to $r$ and this transition is more abrupt compared with other solutions with $q/q_0 \neq 0$ due to the fact that $c_{\chi}^{(-1)} = 0$ for $q/q_0 = 0$.}
\end{figure}

\begin{figure}[h]
 \centering
 \includegraphics[width=15cm]{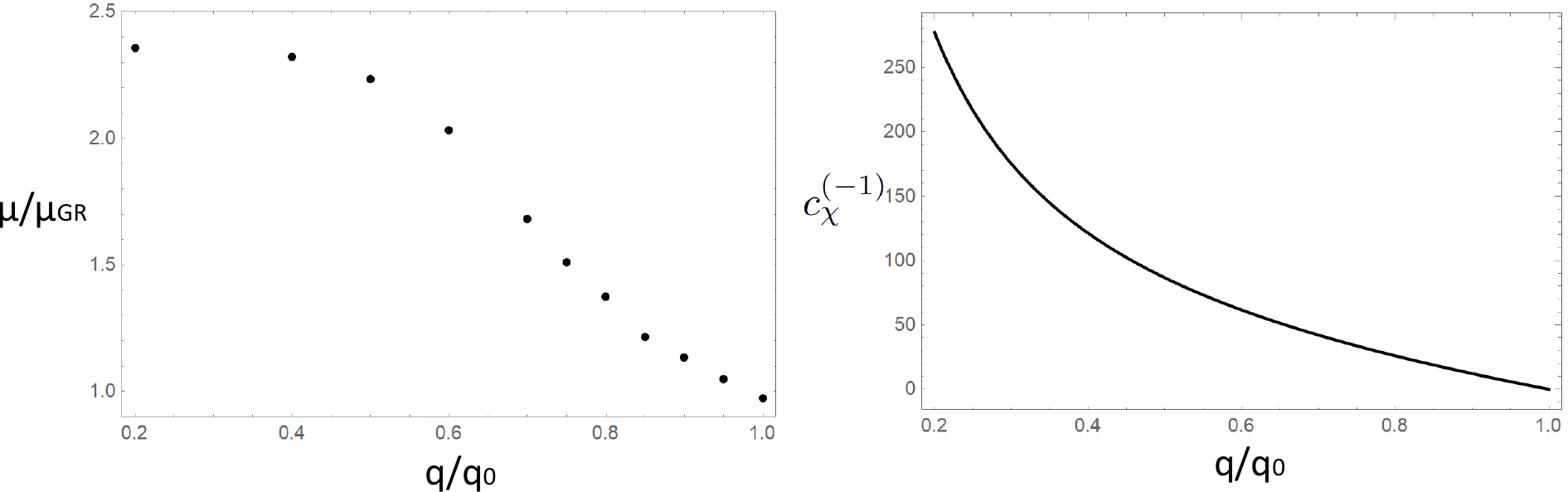}
 \label{Fig3}
 	\caption{Left: The dependence of the mass $\mu$ compared with GR on $q/q_0$. Right: the coefficient $c_{\chi}^{(-1)}$ as a function of $q/q_0$. }
\end{figure}

We should note that our choice of parameters does not represent a realistic situation in our Universe. Numerically, it is a challenge to model a realistic star due the the huge hierarchy between the cosmological horizon, the Vainshtein radius and the size of the star. However, we expect that the qualitative features of the solution remain the same. The deviation from the cosmological solution is determined by $q$, which is an integration constant. Even if there is a huge hierarchy between the Vainshtein radius and the cosmological horizon, the asymptotic solutions (\ref{larger1})-(\ref{larger3}) are valid above the Vainshtein radius thus the deviation from the cosmological solution does not change as long as the solution exists. Note that even if $q \neq q_0$, the scalar field is homogeneous in the Friedman coordinates and the homogenous solution agrees with the cosmological solution even though it contains a slowly decaying inhomogeneous part determined by $q-q_0$ as shown in \cite{Babichev:2016fbg}. On the other hand, we cannot prove the existence of the solution for $q \neq q_0$ for a realistic choice of parameters. We leave this for a future investigation. 

\section{Conclusion}
In this paper, we derived relativistic star solutions numerically in a cubic Galileon theory. In this theory, the Vainshtein mechanism suppresses the scalar field inside the Vainshtein radius. We confirmed that as long as the radius of the star is shorter than the Vainshtein radius, which is the case for a realistic choice of parameters, the stellar structure is indistinguishable from that of GR with the same central density. On the other hand, the scalar field is not suppressed beyond the Vainshtein radius and the scalar field has a charge determined by the mass and the velocity of the scalar field. The velocity of the scalar field can be different from that of the cosmological solution, which is fixed by model parameters. Thus we may interpret this quantity as a scalar hair. We found that the mass of the star determined by the asymptotic behaviour of the metric also depends on the velocity of the scalar field. 

There are several open questions. It is not clear whether the solutions presented in this paper can be formed dynamically by a matter collapse. Also the stability of the solutions needs to be studied. It is also an interesting question whether there is any way to probe the existence of the scalar hair from observations. 

%We also encountered the same difficulty as Ref.~\cite{Babichev:2016fbg} in finding numerical solutions with $q$ that is too different from $q_0$. It would be interesting to establish whether these solutions do not exit and, if so, what determines the critical value of $q$. 

\section*{Acknowledgement}
We would like to thank Antoine Lehébel and Eugeny Babichev for useful discussions. HO was supported by the JSPS overseas challenge program for young researchers and the University of Portsmouth. TK was supported by MEXT KAKENHI Grant Nos.~JP15H05888, JP17H06359, JP16K17707, and JP18H04355. KK is supported by the European Research Council through 646702 (CosTesGrav) and the UK Science and Technologies Facilities Council grants ST/N000668/1 and ST/S000550/1. 

\bibliographystyle{unsrt}
\bibliography{sample}

\end{document}